\documentclass[12pt,preprint]{aastex}

\newcommand{\myemail}{krimm@milkyway.gsfc.nasa.gov}

\slugcomment{Submitted to ApJ}

\shorttitle{GRB 050717}
\shortauthors{Krimm et al.}

\begin{document}


\title{GRB 050717: A Long, Short-Lag, High Peak Energy Burst Observed by Swift and Konus}


\author{H. A. Krimm\altaffilmark{1,2}, C. Hurkett\altaffilmark{3}, V. Pal'shin\altaffilmark{4}, 
J. P. Norris\altaffilmark{1},  B. Zhang\altaffilmark{5}, S. D. Barthelmy\altaffilmark{1}, D. N. Burrows\altaffilmark{6}, N. Gehrels\altaffilmark{1}, 
S. Golenetskii\altaffilmark{4}, J. P. Osborne\altaffilmark{3}, 
A. M. Parsons\altaffilmark{1}, M. Perri\altaffilmark{7}, R. Willingale\altaffilmark{3}}

\email{\myemail}




\altaffiltext{1}{NASA Goddard Space Flight Center, Greenbelt, Maryland, 20771, USA}
\altaffiltext{2}{Universities Space Research Association, 10211 Wincopin Circle, Suite 500, Columbia, Maryland  21044-3432,  USA}
\altaffiltext{3}{Department of Physics and Astronomy, University of Leicester, Leicester, LE1 7RH, UK}
\altaffiltext{4}{Ioffe Physico-Technical Institute, Laboratory for Experimental Astrophysics, 26 Polytekhnicheskaya, St Petersburg 194021, Russian Federation}
\altaffiltext{5}{Physics Department, University of Nevada, Las Vegas, NV  89154, USA}
\altaffiltext{6}{Department of Astronomy and Astrophysics, 525 Davey Lab., Pennsylvania State University, University Park, PA  16802, USA}
\altaffiltext{7}{ASI Science Data Center, Via Galileo Galilei, I-00044 Frascati, Italy}


\begin{abstract}
\noindent The long burst GRB 050717 was observed simultaneously by the Burst Alert Telescope (BAT) on Swift and the Konus instrument on Wind.  Significant hard to soft spectral evolution was seen. Early gamma-ray and X-ray emission was detected by both BAT and the X-Ray Telescope (XRT) on Swift.  The XRT continued to observe the burst for 7.1 days and detect it for 1.4 days.  The X-ray light curve showed a classic decay pattern; the afterglow was too faint for a jet break to be detected.  No optical, infrared or ultraviolet counterpart was discovered despite deep searches within 14 hours of the burst.  Two particular features of the prompt emission make GRB~050717 a very unusual burst. First, the peak of the  $\nu\times F(\nu)$\ spectrum was observed to be $2401_{-568}^{+781}$ keV for the main peak, which is the highest value of $E_{peak}$\ ever observed.  Secondly, the spectral lag for GRB 050717 was determined to be 2.5 {$\pm$} 2.6 ms, consistent with zero and unusually short for a long burst.  This lag measurement suggests that this burst has a high intrinsic luminosity and hence is at high redshift ($z   > 2.7$).   Despite these unusual features GRB 050717 exhibits the classic prompt and afterglow behaviour of a 
gamma-ray burst.
\end{abstract}





\section{Introduction}

\noindent It has long been realized that the full understanding of gamma-ray bursts (GRBs) requires multi-wavelength observations as close together in time as possible.  The unique capabilities of the Swift Gamma-Ray Burst Explorer \citep{gehr04} allow such observations to be carried out rapidly and with high sensitivity at X-ray energies ranging from 0.3~keV to $\sim$350~keV.  When a GRB is also detected simultaneously with the Konus instrument  \citep{apt95} on Wind, one also obtains 
spectral and temporal data up to $>$10~MeV, providing a complete picture of the prompt emission over nearly two orders of magnitude in energy.

When a spectroscopic redshift is not available, it is possible to use features of the prompt emission to constrain estimates of the burst redshift.   In particular, \citet{norris00} noted that pulse peaks migrate to later times as they become wider at low energies.  This spectral lag was found to be proportional to the total peak luminosity of the burst, and can be used along with the peak flux and the peak of the $\nu\times F(\nu)$\ spectrum, or $E_{peak}$ to constrain the absolute luminosity and hence the redshift of the burst.   Also \citet{amat02} and \citet{yone04} have shown that $E_{peak}$\ when converted to the GRB rest frame is proportional to isotropic energy \citep{amat02} or peak luminosity \citep{yone04}. 

After a few hundred  seconds, the prompt gamma-ray emission has decayed and the spectrum has softened to the point where high energy photons are no longer detectable.  However, with a sensitive instrument such as the X-Ray Telescope on Swift, this  late phase can often be detected in X-rays for many days after the initial burst. 
Several authors \citep{zhang05, Nousek_2005, panait05} have presented a unified picture of the time evolution of the early X-ray emission.  In this unified picture, the initial decay component has a steep time decay function where the emission is dominated by the tail of the internal shock  emission \citep{kumar00}, followed by a shallower component where the fireball has decelerated and emission is dominated by the forward shock \citep{mesz97, sari98}. 

The long, bright GRB 050717 was detected by both the Burst Alert Telescope (BAT) \citep{barth05} on Swift \citep{gcn3633} and the Konus instrument on Wind \citep{gcn3640}, allowing simultaneous observations from 14~keV to 14~MeV.  The burst was long enough that it was still detectable in Swift-BAT for $>$60 seconds after it became visible to the Swift X-Ray Telescope (XRT).  The XRT continued to observe the afterglow until 7.1 days after the trigger and it was detectable out to 1.4 days.   No optical transient was found in spite of deep long wavelength searches within 14 hours of the GRB.

In this paper we describe the prompt and afterglow properties of GRB 050717, starting with a description of the Swift, Konus and various optical follow-up observations in Section~\ref{sec2}, and continuing in Section~\ref{sec3} with a discussion of the light curves and spectroscopy from the prompt through the late post-burst phase.  In Section~\ref{sec4} we discuss the implications of these observations, and in particular note the extremely high value of $E_{peak}$\ and unusually short value of the spectral lag.

We shall show that while GRB 050717 is a classical long GRB, based on its spectral and temporal properties, it exhibits several highly unusual and noteworthy features which may constrain burst models.


\section{Observations and Data Analysis}\label{sec2}

\subsection{Swift-BAT}

\noindent At 10:30:52.21 UT, 17 July, 2005, the Swift BAT triggered and located on-board GRB 050717 (BAT trigger 146372) \citep{gcn3633}. Unless otherwise specified, times  in this article are referenced to the BAT trigger time, (UT 10:30:52.21) hereafter designated T$_0$.  The burst was detected in the part of the BAT field of view that was 55\% coded, meaning that it was 36$\degr$ off-axis and only 55\% of the BAT detectors were illuminated by the source.  The spacecraft began to slew to the source location 8.66 seconds after the trigger and was settled at the source location at T$_0$+63.46 seconds.

The BAT data for GRB 050717 between T$_0$-300 s and T$_0$+300 s were collected in event mode with 100 $\mu$s time resolution and $\sim$6~keV energy resolution.   The data were processed using standard Swift-BAT analysis tools and the spectra were fit using  {\sc xspec 11.3}.  Each BAT event was mask-tagged using {\sc batmaskwtevt} with the best fit source position. Mask-tagging is a technique in which each event is weighted by a factor representing the fractional exposure to the source through the BAT coded aperture.  A weight of +1 corresponds to a fully open detector and a weight of -1 to a fully blocked detector.  Flux from the background and other sources averages to zero with this method.  All of the BAT GRB light curves shown have been background subtracted by this method.  This method is effective even when the spacecraft is moving since complete aspect information is available during the maneuver. 

The mask-weighting is also applied to produce weighted, background subtracted counts spectra using the tool {\sc batbinevt}.  Since the response matrix depends on the position of the source in the BAT field of view, separate matrices are derived for before the slew, after the slew and for individual segments of the light curve during the slew. 

\subsection{Konus-WIND}

\noindent The long hard GRB 050717 triggered Konus-Wind (K-W) \citep{apt95}  at 
T$_0$(K-W) = 10:30:57.426 UT.  
It was detected by the S1 detector which observes the south ecliptic
hemisphere; the incident angle was $55\fdg5$. 
The propagation delay from Swift to Wind is 2.369~s for this GRB,  {\em i.e.}, correcting for this factor,
one sees that the K-W trigger time corresponds to T$_0$+2.86~s.
The data before T$_0$(K-W)-0.512~s were collected in the waiting
mode with 2.944~s time resolution.   From T$_0$(K-W) to T$_0$(K-W)+430.848~s, 64 spectra in 101
channels were accumulated.
The first 4 spectra were accumulated on a 64-ms time scale, then
the spectra accumulation times were varied from 5.120~s to 8.192~s
adapting to the current burst intensity. Data were processed using standard Konus-Wind analysis tools
and the spectra were fitted by {\sc xspec 11.3}.  As observed by Konus-Wind GRB 050717 had a steep rise and a long decaying tail. 

\subsection{Swift-XRT}

\noindent The spacecraft slewed immediately to the BAT location of GRB 050717 and the XRT began observing the burst at 10:32:11.49 UT (approximately 79 seconds after the BAT trigger). The automated on-board XRT software was unable to centroid on the burst, however, the downlinked X-ray spectrum and light curve clearly showed a bright fading X-ray object in the field. XRT observations \citep{gcn3636}  began in Windowed Timing (WT) mode (see below) 91 seconds after the trigger before going into Photon Counting (PC) mode at 310~s. The coordinates of the burst were determined by the XRT to be (J2000): RA:$14^\mathrm{h} 17^\mathrm{m} 24\fs58$ ($214\fdg 352$),
   Dec: $-50\arcdeg 31\arcsec 59\farcs92$ ($-50\fdg533$)
(the 90\% confidence error circle radius is 3.5 arc seconds) \citep{moretti05}.  

Swift's X-ray Telescope uses a grazing incidence Wolter I telescope to focus 
X rays onto a CCD-22 detector. 
It has an effective area of 135 cm$^{2}$ at 1.5 keV and an angular resolution of 18 arcsec. 
For further information on the 
XRT see {\em e.g} \citep{Burrows_2003, gehr04, hill04, burro05}. This instrument has three key functions:
the rapid, automated and accurate determination of GRB positions, the provision 
of moderate resolution spectroscopy 
(energy resolution 140 eV at 5.9 keV), and recording GRB light curves over a wide 
dynamic range covering more  than seven orders of magnitude in flux. 

The WT readout mode of the XRT uses a restricted 
portion of the telescope's total field of view: the central 8 arcmin (or 200 columns), 
when the GRB flux is below $\sim$5000~mCrab. 
Each column is clocked continously to provide timing information with 
1.8~ms resolution. However, this rapid readout 
mode only preserves imaging information in one dimension. 
Once the GRB flux drops below $\sim$1~mCrab the PC mode takes over. 
This mode retains full imaging and spectroscopic information with a readout time of 2.5~s.

Data for this burst were obtained from the Swift Quick Look website~\footnote{http://swift.gsfc.nasa.gov/cgi-bin/sdc/ql?} 
and processed with version 2 of the Swift software. The {\sc xselect} program
was used to extract source and background spectra and 
cleaned event lists (0.3--10.0~keV), using  {\sc xselect} grades 0--12 for PC mode data and grades 0--2 for WT data.

The PC mode suffers from pile-up when the count rate is $\geq$~0.8 counts s$^{-1}$. 
To counter this we extracted a series 
of grade 0--12 background corrected spectra from the first 8.6~ks of PC mode data using 
annuli of varying inner radii. We 
deem the point at which pile-up no longer affects our results to be when the spectral 
shape no longer varies with an 
increase in annular radius. For GRB 050717 this occurred when we excluded the inner 
12 pixels (radius). Only the first 500~s 
of PC mode data suffered from pile-up. The WT data were free from pile up problems. 
The spectra were then analyzed as normal 
in {\sc xspec 11.3.21}. The light curve was created by the same method as 
detailed in \citet{Nousek_2005}.

\subsection{Swift-UVOT} \label{uvot}

\noindent Observations with the Swift Ultra Violet/Optical Telescope (UVOT) began at 10:32:10.7 UT (78 seconds after the BAT trigger) \citep{gcn3638} . The first datum taken after the spacecraft settled was a 100~s exposure using the V filter with the midpoint of the observation at 128 s after the BAT trigger. No new source was detected within the XRT error circle  in summed images in any of the six filters down to the 
3$\sigma$ magnitude upper limits shown in Table~\ref{tbl-1}.

\subsection{Other observations} \label{other}

\noindent GRB 050717 was  not well positioned for follow-up observations.  Its high southern declination made it unobservable by most northern hemisphere telescopes and the trigger was just before dawn at the South American observatories.  Consequently, no  follow-up optical observations were made until more than 13 hours after the burst.  In the several observations that were made after this time, no optical counterpart was detected.

Under the control of Skynet, the Panchromatic Robotic Optical Monitoring and Polarimetry Telescopes (PROMPT) automatically observed the refined  XRT localization of GRB 050717 beginning 13.0 hours after the burst  \citep{gcn3652}.  No source was detected within this localization. Limiting magnitudes ($3\sigma$), based on 5 USNO-B1.0 stars, are 21.7 (R$_c$, T$_0$+13.67~hr) and 21.5 (I$_c$, T$_0$+16.02 hr). 

Observations in the K-band were made with the Wide-Field Infrared Camera on the du~Pont 100-inch telescope at Las Campanas Observatory on two occasions: 2005 July 18.01 UT (T$_0$+13.7 hr) \citep{gcn3639} and on 2005 July 18.98 UT (T$_0$+37.0 hr) \citep{gcn3643}.  Within the $6'$ radius XRT error circle four sources were found of which one is also visible in 2MASS K-band images. The other three sources have magnitudes of 18.1, 18.7, and 19.2 in comparison to several 2MASS stars; the $3\sigma$\ limiting magnitude of the image is about 19.4.  None of the three uncataloged objects faded between the two observations.  In addition, \citet{gcn3643} obtained I-band images with the LDSS-3 instrument on the Magellan/Clay telescope on 2005 July 18.06 and 18.97 UT (14.9 and 36.8 hours after the burst, respectively). The same three sources visible in the K-band images were detected but had not faded.

\citet{gcn3642} obtained six five-minute unfiltered images on 2005 July 18.46 UT (24.5 hours after the burst), using one of Tenagra observatory's 0.35-m telescopes with an AP6 CCD at Perth, Western Australia. No new source was detected within the XRT error circle of GRB 050717 down to the DSS-2R limiting magnitude.

\section{Light Curves and Spectroscopy}\label{sec3}





\subsection{Swift-BAT} \label{bat}

\noindent The BAT triggered on the first of two short, small spikes that preceded the main emission of GRB 050717.  This first spike at T$_0$\ was very soft (photon power-law spectral index 2.89 $\pm$ 0.14) and lasted 128 ms.  The second short spike began at  T$_0$+0.7 s, was of longer duration (320 ms) and was much harder (photon index 1.36 $\pm$ 0.23).  The precursors are shown in detail in the left-hand panels of Fig~\ref{fig1}. These small precursors were followed by the main pulse, which displayed the common fast rise, exponential decay (FRED) profile.  The intensity rose from background to peak within 450 ms then began to decay with an average exponential decay constant $1.82^{+0.13}_{-0.11}$.  The full light curve is shown in the right-hand panels of Fig~\ref{fig1}. The peak count rate was measured by BAT to be $\sim$16000 counts s$^{-1}$ at T$_0$+4 s in the 15-350 keV band.  On top of this slow decay, there were at least four other peaks, showing a gradual spectral softening.    The duration T$_{90}$ (15--350 keV) is 86 $\pm$ 2 s (estimated error including systematics). The total fluence in the 15--350 keV band is (1.40 $\pm$ 0.03) $\times 10^{-5}$ erg cm$^{-2}$. The 1-s peak photon flux measured from T$_0$+2.8 s in the 15--350 band is 8.5 $\pm$ 0.4 ph cm$^{-2}$ s$^{-1}$. All the quoted errors are at the 90\% confidence level.   The fluence hardness ratio for this burst is S(100-300~keV)/S(50-100~keV) =   $(8.13 \pm 0.14)\times 10^6$\ erg/cm$^2$\ /   $(2.23 \pm 0.06)\times 10^6$\ erg/cm$^2$\ = 3.65

The BAT data were binned into eleven time bins to track the spectral evolution of the prompt emission.  This is shown in the lower panels of Fig~\ref{fig1}.  Starting with the main peak, there is clear evidence of spectral softening as the burst progresses.  Then after T$_0$+91 s, the BAT spectrum hardens again.  The  fit to the BAT data only over T$_0$+91 to T$_0$+150 s yields a power-law photon index of 1.08 $\pm$0.32.  A joint fit to the BAT and XRT data  over the same time period (see  Section~\ref{xrt}) gives a photon index of 1.61 $\pm$ 0.08.  The low BAT flux at these times limits statistically meaningful fits to the entire interval.  However, given the spectral variation demonstrated earlier in the burst, it is quite possible that there is spectral evolution occurring at these times as well and the overall spectral fits should be interpreted with caution.

\subsection{Konus-WIND} \label{wind}

\noindent 
The Konus-Wind light curve is shown in three energy bands in Fig~\ref{fig2}, and the 21--1300 keV  light curve (see Fig~\ref{fig2a}) is similar to the Swift-BAT light curve.  The long decaying tail is clearly seen in G1 band (21--84~keV),
marginally seen in G2 band (84--360~keV), and not seen in G3 band (360--1370~keV).
The G2/G1 ratio demonstrates substantial softening of the tail as compared
to the main pulse. The T$_{90}$ durations of the burst in G1, G2, G3
energy bands are 99$\pm$10~s, 95$\pm$11~s, 18$\pm$3~s, respectively.  For the sum G1+G2+G3, the T$_{90}$\ duration is 96$\pm$6~s.

Emission is seen up to $\sim$10~MeV.   We were able to fit the data 
in the 20 keV--6 MeV range by a power law model with an exponential cutoff:  $ F(E)\ =\ A \times (E/100\ {\rm keV})^{-\alpha} \times  \exp\left(\frac{-E(2-\alpha)}{E_{peak}}\right) $, where $E$\ is the energy in keV, $E_{peak}$\ is the peak energy of the $\nu\times F(\nu)$\ spectrum, $\alpha$\ is the photon index, and $A$\ is a normalization factor.    For the time integrated spectrum (T$_0$+2.843~s to T$_0$+54.555~s) we find $\alpha = 1.19 \pm 0.12$\ and $E_{peak}$ = $2101_{-830}^{+1934}$ keV ($\chi^2$= 88  for 77 d.o.f.). The spectrum of the main peak (from T$_0$+2.843~s to T$_0$+8.219~s) is well fitted with $\alpha = 1.05 \pm 0.10$\ and $E_{peak}$ = $2250_{-620}^{+940}$ keV  ($\chi^2$= 83  for 85 d.o.f.). Fitting jointly with the BAT data for the main peak gives  $\alpha = 1.04 \pm 0.05$\ and $E_{peak}$ = $2401_{-568}^{+781}$ keV ($\chi^2$= 117  for 143 d.o.f.).  These values of $E_{peak}$\ for both the time integrated and time resolved spectra are perhaps the largest ever measured.  The implications of this are discussed in Section~\ref{epeak}. Figure~\ref{fig1a} shows that the BAT and Konus data can be well fit to the same model spectrum.  A fit to the Band (GRBM) model was also attempted.  No statistically significant high energy power-law tail was established.  The limit on the high energy photon index is $\beta > 1.89$\ (90\% C.L.)  The low energy photon index $\alpha$\ is almost the same as for the cut-off power-law model, $\alpha = 1.02_{-0.3}^{+0.7}$ .  

	Joint fits between BAT and Konus were also made for two later time intervals:  T$_0$+13.851~s -- T$_0$+26.907~s, and T$_0$+26.907~s -- T$_0$+54.555~s.  The photon indices for a simple power-law fit are shown in Figure~\ref{fig1}.   The first of these intervals was  also fit with a cut-off power law, but only a lower limit to $E_{peak}$ was found:  $E_{peak} > 548$\ keV (90\% C.L.).  We were unable to  make a well-constrained joint BAT-Konus fit to the full burst due to problems creating a single response matrix to cover both the slew and the period before the slew.  

The total fluence in the 20~keV to 6~MeV range is  $6.5_{-2.2}^{+0.9} \times 10^{-5}$ erg cm$^{-2}$. The 64-ms peak flux measured from T$_0$+2.86 s in the same energy band is $1.41_{-0.24}^{+0.18} \times 10^{-5}$ erg cm$^{-2}$ s$^{-1}$. The uncertainties in the derived fluence and peak flux are dominated by uncertainties in the high energy part of the spectrum.   

All quoted uncertainties are at the 90\% confidence level. 

\subsection{Swift-XRT} \label{xrt}

\noindent The spectrum between 91 s and 310 s after the trigger (WT data) has an average photon index of $1.65\pm 0.11$, with the absorption fixed at its galactic value of $2.22 \times 10^{21}$ cm$^{-2}$ and an indication of an excess absorption of $2.75\pm 0.57 \times 10^{21}$ cm$^{-2}$\ , assuming $z=0$ and standard (local) interstellar material abundances. The mean unabsorbed flux in WT mode at 201 s (mean time) is 
$5.76\pm 0.31 \times 10^{-10}$ erg cm$^{-2}$ s$^{-1}$ in the 0.3-10.0 keV energy range.   

During the period between T$_0$+91 s and T$_0$+150 s, a joint fit was made to the XRT and BAT data.  The joint fit gives a photon index of $1.61 \pm 0.08$, with an excess absorption of $3.36^{+0.8}_{-0.68} \times 10^{21}$ cm$^{-2}\ (\chi^2=125$\ for 115 d.o.f.).   This fit was used to extrapolate the BAT 15-150 keV flux into the XRT energy range (0.3--10 keV) during the overlap interval assuming that the 1.61 power law index holds in both energy ranges.  Since we know the BAT count rate in the BAT (15--150 keV) range, we were able to use {\sc xspec} to derive the model flux in the 0.3-10 keV band and then calculate a ratio between BAT counts (15--150 keV) and flux (0.3--10 keV).  For earlier epochs we derived the conversion ratio from the model fits to the BAT data alone.  We derived a similar ratio between XRT counts (0.3--10 keV) and flux.  With this extrapolation one can directly compare the early and later light curves  and show (Figures~\ref{fig3} and~\ref{fig4}) that the prompt emission smoothly transitions to the afterglow emission.

The data from T$_{0}$+1.17 hr to T$_{0}$+8.25 hr were also fit with a
power law with a photon index of 1.35$\pm$0.21 and galactic absorption
($\chi^{2}$=16.9 for 11 d.o.f.). The model flux over 0.3--10.0 keV was
1.8$\pm$0.41 $\times10^{-12}$ ergs cm$^{-2}$ s$^{-1}$
(3.54$^{+0.89}_{-1.00}$ $\times10^{-4}$ photons cm$^{-2}$ s$^{-1}$).
 In this case there was no improvement to the fit by adding excess absorption.  Indeed, this later
 spectrum is not consistent with excess absorption at the level implied by the earlier WT data; the
 excess absorption is limited at 90\% confidence to $< 1.5 \times 10^{21}$\ cm$^{-2}$.  

\subsection{Post-Burst Emission} \label{after}

\noindent The gamma-ray and X-ray decay light curve  is shown in Fig~\ref{fig3}. The light curve shows several prominent features which can be interpreted in light of the models discussed in \citet{zhang05} (hereafter Z05).  First, as pointed out earlier there is a smooth transition  from the prompt BAT emission into the early X-ray emission and a fairly steep decay (power law index $\alpha_1$\ in the discussion below) until T$_0 >\ \sim200$\ s.  This is followed by a possible superimposed X-ray flare, a phenomenon quite common in GRBs as observed by Swift \citep{burro05b, barth05b}. Unfortunately  observing constraints cut off observations in the middle of the possible flare, and the statistics do not allow for a meaningful fit to a flare component.   Observations resumed again at T$_0$+ 4214~s, with a return to a power-law decay, with a shallower power law index ($\alpha_2$\ below). 

In order to fit the data to reasonable X-ray emission models, two intervals were removed:  BAT data points before T$_0$+ 50~s, which were believed to be part of the prompt emission, and XRT data points between T$_0$+500 and the end of the first observation, so that the fit is not contaminated by the possible flare.  Two different fits were made and are discussed in turn.

First, we tried a broken power law.  This gave a power law index $\alpha_1 = 2.10 ^{+0.17}_{-0.05}$\ for the steep part of the light curve,  a break time of $203 \pm 26$~s and an index $\alpha_2 = 1.48 \pm 0.02$\ for the shallow part ($\chi^2 = 159$ for 111 d.o.f.).    The  steep part of the curve ($\alpha_1 = 2.10$) corresponds to region I in Fig.~1 of Z05.  According to Z05, if this time can be interpreted as the curvature effect \citep{kumar00,dermer04}, the index should be $\alpha = 2 + \beta$, where $\beta$ is the {\em energy} index of the spectrum of the emission.  Taking $\beta$ = 0.62, we should have $\alpha_1$ = 2.62, as compared to the observed value of  2.10. 

One factor which could lead to a deviation from the $\alpha = 2 + \beta$\ relation is that the decay curve seen could be a superposition of two separate decay power laws, one steep due to the curvature component, and one shallow due to the forward shock component.    So a fit was made to a superposition model:  $F(t) = A \times t^{-\alpha_1} + C \times t^{-\alpha_2}$, where $A$\ and $C$\ are normalization factors.  This fit gave a steep index $\alpha_1 = 3.01^{+0.55}_{-0.23}$\ and a shallow index $\alpha_2 = 1.43\pm 0.04$ ($\chi^2 = 161$ for 110 d.o.f).   Statistically this fit is indistinguishable from the broken power law.  However the physical interpretation is more straightforward.  The steep index ($\alpha_1 = 3.01$) is the decay of the tail of the internal shock emission, which is superimposed on an underlying afterglow component with a decay index of $\alpha_2 = 1.43$.  The afterglow component becomes dominant at T$_0+\ \sim100$~s.

It is instructive to compare the measured temporal index ($\alpha_2 \approx 1.4$) with the values predicted by the simple afterglow models compiled by \citet{zhang04}.  At late times (t  $> $1.17 hr), we should be in the slow cooling regime, and the spectral index of GRB~050717, $\beta = 0.35 \pm 0.21$, is consistent only with the regime where $\nu_m < \nu < \nu_c$. Here, following  \citet{zhang04}, $\nu$\ is the spectral frequency of the emission, and $\nu_m$\ and $\nu_c$ are the synchotron frequency and cooling frequency, respectively.  Using $\beta = 0.35 \pm 0.21$, we have the electron-acceleration power-law index  $p = 1 + 2\beta = 1.7 \pm 0.4$.  Using the equations in Table 1 of \citet{zhang04}~\footnote{We have changed the signs of $\alpha$ and $\beta$ in the equations of \citet{zhang04} to conform to the definition F$_{\nu} \propto t^{-\alpha} t^{-\beta}$\ used in this paper.} and taking $p > 2$, we derive values for $\alpha$ of $0.5 \pm 0.3$, and $1.0 \pm 0.3$ for the ambient interstellar medium (ISM) \citep{mesz97, sari98} and wind  models \citep{chev00}, respectively.  If $1 < p < 2$ \citep{dai01}, we derive $\alpha$ values of $0.7 \pm 0.1$, $1.2 \pm 0.05$, again for the ISM and wind models, respectively.  We see that the late-time temporal index ($\alpha_2 \approx 1.4$) is inconsistent with the ISM model and marginally consistent with the wind model. This analysis shows that at late times emission is dominated by the forward shock with a wind density profile.  

In order for the afterglow of GRB~050717 to have $\nu_m < \nu < \nu_c$, it must be observed at a time such that t $>$\ t$_c$, where the critical time t$_c$\ is defined in Z05.  This puts constraints on the wind parameter A$_*$, which is defined in \citet{chev00} as being proportional to the wind mass loss rate divided by the wind velocity (units g cm$^{-1}$).  The parameter A$_*$\ must be in the range 0.01--0.001, which is similar to the limit derived for GRB~050128 \citep{camp05}.

The late time shallow decay (index $\sim 1.4$)  continues until the flux becomes unobservable to the XRT.  A lower limit is set for summed observations after T$_0$+2.6~days.  Since there is no apparent break to a steeper decay in the light curve, the lower limit on a jet-break time is $t_b > 1.4$~days.

\section{Discussion}\label{sec4}

\subsection{Spectral Lag} \label{prompt}

\noindent It is possible to derive an estimate of the spectral lag of the BAT data between Channel 2 
(25-50 keV) and Channel 4 (100--350 keV).  From the spectral lag we can use the methodology of \citet{norris00} and \citet{norris02} to derive limits on the redshift of the GRB and on the isotropic luminosity of the peak of the emission.   The spectral lag was derived for the main peak of emission (from T$_0$+2.26~s to T$_0$+5.8~s).  The lag was found to be $2.5_{-2.4}^{+2.9}$~ms.  Hence the measured lag is statistically consistent with zero.  
The lag was also measured for several other intervals during the burst and with time rebinning ranging from 2~ms to 16~ms.  In all  cases, the measured lag was small, positive and consistent with zero.
Such a low value for lag is quite unique for a long burst since \citet{norris02} has shown that the dynamic range of lags for long bursts spans $\sim25$\ ms to $\sim300$~ms.   In fact out of the 90 brightest bursts studied by \citet{norris02}, only 2\% show a lag as small as that of GRB 050717.

One can use the lag, the measured peak flux, and $E_{peak}$ to set lower limits on the distance to the burst.  Using the peak flux of $1.69\pm 0.16 \times 10^{-6} $ ergs cm$^{-2}$ s$^{-1}$ (15--350~keV;  T$_0$+2.752~s to T$_0$+3.008~s), the parameters from the joint Konus-BAT fits to the main peak (Section~\ref{wind}) and the $+2\sigma$ limit on the lag (8.3 ms), one derives a redshift of 2.7 and a peak luminosity of $3.9 \times 10^{53}$ ergs s$^{-1}$ (15--350 keV).  The fit is relatively insensitive to variations in either peak flux or $E_{peak}$ and other spectral fit parameters.  Since smaller values of spectral lag would lead to larger redshifts, this value, z = 2.7, can be considered the $2\sigma$ lower limit on the redshift; similarly the luminosity is also a lower limit.  Such a large redshift is consistent with the non-detection of an optical or infrared counterpart to the afterglow (Sections~\ref{other} and \ref{optical}) and with the non-detection of a jet break (Section~\ref{after}).  

A consistent interpretation of such a small lag is that the high energy emission from GRB~050717 has been redshifted downward more than usual into the BAT energy range.  It has been shown \citep{norris96, fenim95} that the high energy component of burst emission shows narrower peaks and more variation than is seen at lower energies.  Shifting such spiky peaks into the BAT range would cause the measured lag to be smaller than what would be observed in long bursts at lower redshifts.  

\citet{norris05} have pointed out that many short bursts seen by BATSE, Swift, Konus-Wind, and HETE-2 have extended emission starting a few seconds after the short spike and lasting for $\sim$tens of seconds.  Since short bursts are also known  to have short lags \citep{norris01}, is it possible that GRB~050717 is in fact a short burst or a magnetar flare from a nearby galaxy?  This burst has a pair of precursors of duration 128~ms and 320~ms, followed by $> 100$~s of extended emission along with a spectral lag consistent with short GRBs.  However, two properties of GRB~050717 argue strongly against it being a short burst.  First of all, the spectra of the precursors of this burst are significantly softer than the extended emission (see Fig~\ref{fig1}), while in all short bursts with extended emission the short spikes are significantly harder than the extended emission.  Secondly, in GRB~050717, the flux is dominated by the extended emission, while in short bursts, the flux is dominated by the short episode of emission.  The ratio of peak to tail emission for a magnetar \citep{palm05} is even more extreme. Therefore, it is more likely that GRB~050717 is indeed a long burst seen at a large distance.
	
Using a relationship derived by \citet{liang05} we can use the measured $E_{peak}$ and the limits on luminosity and redshift to set a lower limit on the jet break time for this burst.   After re-arranging Equation~5 in \citet{liang05}:

\begin{equation}
t_b = 0.88 \times (E_{\gamma,iso,52})^{-0.81} \times \left(\frac{E_p}{100 keV}\right)^{1.56} \times (1+z)^{2.56}.
\end{equation}\label{eq2}

\noindent Here $t_b$ is the jet break time in days in the observer frame, $E_{\gamma,iso,52} >100$  is the isotropic energy in units of $10^{52}$~ergs and $E_p$=2400~keV is the observed peak energy.  Errors on the exponents in the equation have been suppressed since the calculation is dominated by errors in the input parameters.  With these values and z=2.7, we can derive a lower limit on $t_b$ of 88 days.  As we saw in Section~\ref{after} this is fully consistent with the observations. 

\subsection{Comparison to Other GRBs} \label{epeak}

\noindent It was noted in Section~\ref{wind} that $E_{peak}$\ for GRB~050717 is unusually high.  The observed values are 2400~keV for the main peak and 2100~keV for the time integrated spectrum;  when propagated to a rest frame at $z$\ = 2.7, the intrinsic values of $E_{peak}$\ become 8900~keV and 7800~keV, respectively.  These values can be compared to previously measured values of $E_{peak}$\ from the Burst and Transient Source Experiment (BATSE) and BeppoSAX. \citet{kane06a} (see also \citet{kane06}) have performed a systematic spectral analysis of 350 bright GRBs observed by BATSE.  Of these bursts, none show an integrated $E_{peak}$\ as large as what was measured for GRB~050717.  The highest value calculated was 2039~keV for GRB~971220.  \citet{pree00} provide time resolved spectroscopy for 156 bright BATSE bursts. In studying the catalog provided with the \citet{pree00} paper, we found only two bursts which had $E_{peak} > 2000$\ keV in multiple time resolved spectra. The moderately bright burst  GRB~940526B had $E_{peak} > 2000$\ keV in seven of the nine time resolved spectra,  although \citet{kane06a} report that the best fit $E_{peak}$\ for this burst is 1689~keV.    One other BATSE burst, GRB~960529, had well constrained values of $E_{peak} > 2000$\ keV in a number of its time resolved spectra, and  an unconstrained time-integrated value of  $E_{peak} > 2000$\ keV.    It is clear from Figure~21 in \citet{kane06} that only a very small fraction of the 8459 time resolved spectra fit by \citet{kane06} have $E_{peak}$\ values as large as what is found for GRB~050717. It should be noted however, that the Konus energy range extends farther than does BATSE ($\sim 2$\ MeV), meaning that some bursts with extremely high values of $E_{peak}$\ may not be well constrained in the \citet{kane06} or \citet{pree00} fits.  In addition none of the twelve BeppoSAX bursts studied in \citet{amat02} have $E_{peak}$\ as high as what we report for GRB~050717.  Clearly GRB~050717 is an exceptional case. 

It is instructive to ask if GRB~050717 is unusual in other ways.  We can compare for example the position of GRB~050717 on a hardness-duration plot to other samples.  Using $T_{90}$ = 86~s and the fluence ratio S(100-300~keV)/S(50-100~keV) = 3.67, we can see that this burst does not have a particularly high hardness ratio and falls well within the scatter of long bursts in both the BATSE and BAT samples (see for example \citet{saka06}).

One can also use the redshift inferred from the spectral lag (Section~\ref{prompt}), to see how GRB~050717 fits the relationships that previous authors have derived between $E_{peak}$\ and isotropic energy \citep{amat02} and peak luminosity \citep{yone04}.  The lower limits on the isotropic radiated energy,  
    the isotropic peak luminosity, and peak energy in the source rest
frame corresponding to the limit z $>$\ 2.7 are         $E_{\gamma}^{iso} > 1.1 \times 10^{54}$~ergs (1-10,000 keV),
    $L_{max}^{iso} > 9.6 \times 10^{53}$~erg~s$^{-1}$ (30-10,000~keV), 
    and $E_{peak}^{rest} > 7800$~keV
    (for a standard cosmology: $\Omega_M = 0.3$, 
    $\Omega_\Lambda = 0.7$, $H_0 = 70$~km~s$^{-1}$~Mpc$^{-1}$).  These values show that GRB~050717 is an outlier on both the \citet{amat02} and \cite{yone04} relations in the direction of  $E_{\gamma}^{iso}$\ and  $L_{max}^{iso}$  being smaller than the relations would predict given $E_{peak}^{rest} = 7800$~keV.  It is not possible to adjust the redshift (within reasonable limits: 
 $z < 20$) to bring GRB~050717 in line with either relation. Thus we must conclude that GRB~050717 does not fit either the \citet{amat02} or \citet{yone04} relations. We note that the maximum value of $E_{peak}^{rest}$\ used in the derivation of either relation is $\sim 2000$\ keV, so these relations have not been verified for values of $E_{peak}$\ as large as that of GRB~050717.
    
There have also been a number of recent papers \citep{band05, nakar05a, nakar05b, kane06} which have presented strong evidence that the Amati relation is not universal and that there are many BATSE GRBs for which the relation is inconsistent.  GRB 050717 is an  excellent example demonstating that the \citet{amat02} relation does not hold for all bursts, particularly those with high $E_{peak}$.   Maybe GRB 050717 and similar bursts will eventually be able to tell us
why the Amati relationship is breaking down.

\subsection{Lack of Optical Counterpart} \label{optical}

\noindent As noted in Sections~\ref{uvot} and \ref{other}, no optical counterpart to GRB 050717 was found.  The deepest limits were those obtained from PROMPT, at 21.7 (R$_c$, T$_0$+13.67~hr) and 21.5 (I$_c$, T$_0$+16.02 hr).  What conclusions can be drawn from the lack of an infrared counterpart? 

First of all, is GRB~050717 a dark burst? \citet{jakob04} make a comparison between the observed X-ray flux and the R-band magnitude of the afterglow at ten hours after the trigger for a large set of bursts and define a dark burst as a burst lying in a certain region of the log(F$_{\rm opt}$)-log(F$_X$) diagram.  For GRB~050717, the X-ray flux interpolated to T$_0$+10~hr is 0.015 $\mu$Jy (see  Figure~\ref{fig3}) and the R-band limit extrapolated to T$_0$+10~hr would be R$_c \approx$21.5.  This is solidly within the bright burst region of the \citet{jakob04} diagram; thus it is not possible to say that this is a dark burst given how late the optical limits are.  

Similarly the lack of a counterpart cannot be used as confirmation of the high redshift.  Assuming $z = 2.7$, the Lyman edge would be redshifted to 91.2~nm($z+1$) = 337~nm.  This is consistent with the relatively shallow ultraviolet limits set by UVOT, but the counterpart could still easily be observed in the I band.  It is instructive to compare the infrared observations of GRB~050717 to those of GRB~050904, a high redshift ($z=6.29$) burst for which an infrared counterpart was found.  However, the IR observations of GRB~050904 were either much earlier (J$\sim 17.5$, T$_0+\sim3$~hr) \citep{gcn3913} or much deeper (I$\sim22.9 \pm 0.6$, T$_0+ \sim 37$~hr) \citep{gcn3932} than those obtained for GRB~050717.  The lack of an observed counterpart to GRB~050717 must be attributed to the lateness of the observations.




\section{Conclusions}

\noindent The long gamma-ray burst GRB~050717 shows a number of interesting features which can be interpreted in light of the predominant models of bursts and their afterglows.     

It has been known for many years that spectral evolution operates in long gamma-ray bursts in several ways.  \citet{golen83} first recognized that the more intense portions of bursts are spectrally harder than the less intense time periods.  Concomitantly, individual burst pulses are asymmetric, especially at low energies.  This was pointed out by \citet{norris96} and later \citet{band97} and \citet{norris02} showed via spectral lag analysis that, if the burst was bright enough, positive lags were manifest, averaged over the whole time profile.  Similarly, and related to the first two effects, the burst ``envelope'' (containing the peaks and valleys in a burst) tends to soften with time in the vast majority of bursts, an effect that was quantified by \citet{band98}.  \citet{nemir94} tied these effects together by demonstrating conclusively that on all time scales, gamma-ray bursts are time-asymmetric.  Thus the later, usually lower intensity portions of a burst should also be spectrally softer.

The long gamma-ray burst GRB~050717 shows all aspects of these evolutionary trends including overall hard to soft spectral evolution as the prompt emission decays and time asymmetries in all peaks at all energies.  Features include two short, soft precursor spikes and at least seven peaks in the main burst.

The main emission of the burst clearly exhibits hard to soft spectral evolution as discussed in \citet{zhang04} and \citet{norris86}.  
The light curve of the prompt emission (Fig~\ref{fig1}) begins with two short, faint, spectrally soft spikes, followed by an intense peak which is the hardest portion of the burst.  The burst intensity envelope as seen above 15 keV decays over the next $\sim 150$\ seconds until it becomes detectable only at lower energies.  Superimposed on the overall decay are at least four subsidiary peaks, each of which is less intense and softer than the one before.  However the spectra of the peaks are harder than the intervening valleys.  Furthermore, as seen in Figs~\ref{fig1} and \ref{fig2a}, each peak is time-asymmetric at all energies.  Thus the time profile of this burst is a very good example of the overall time asymmetry described by \citet{nemir94}.

\citet{norris96} also showed that the structure of pulses in GRBs is narrower at high energies.  This is another aspect of what \citet{norris96} has called the ``pulse paradigm,'' and is physically related to the overall spectral evolution of pulses.  GRB 050717 was unusual in that its spectral lag is very short (positive but statistically consistent with zero -- see Section~\ref{prompt}), while nearly all long bursts clearly show a large positive spectral lag \citep{norris02}.   The short lag and observed brightness of the burst suggest that it is at a high redshift ($z > 2.7$) and hence has a large intrinsic luminosity ($L_{peak} > 9 \times 10^{53}$ erg s$^{-1}$).  The features observed in the burst are likely representative of spiky high energy features red-shifted to the BAT energy range.

 The late decay of GRB 050717
is consistent with a steep decay from the tail of the internal shock emission superimposed on a less steep underlying afterglow component.  At later times after the fireball has decayed, the emission is dominated by the forward shock component with an inferred X-ray flare, followed by a shallow decay. 

GRB 050717 also demonstrates many of the features of the unified picture of the late time evolution of GRB emission \citep{zhang05, Nousek_2005, panait05}. When the BAT flux is extrapolated to the 0.3--10 keV energy range it is seen that the prompt emission smoothly transitions into the slowly decaying phase.  During the early X-ray emission of GRB 050717, the decay index is somewhat less steep than would be expected if it were due solely to the tail emission of the prompt GRB.  As discussed in Section~\ref{after}, this can be interpreted as a superposition of tail and external shock emission, although other interpretations are also discussed.   Before data collection was cut off by an orbital constraint at $\sim$~800 s after the trigger, the light curve shows evidence of the start of an X-ray flare,   When observations take up again, the flux is much weaker and the decay    index is shallow, since at this time the afterglow is dominated by the forward shock. The flux became too faint to observe before the expected jet break at $t_b > 90$ days.

The short spectral lag and high $E_{peak}$\ are very unusual for long GRBs, putting GRB~050717 within the bottom 2\% of long bursts for spectral lag, and within the highest few bursts detected in terms of peak energy. Other burst and afterglow properties are common and easily interpreted.  This is an indication that these properties also hold for bright, high redshift bursts.

\acknowledgments

HAK was supported in this work by the Swift project, funded by NASA. This work is supported at the University of
Leicester by the Particle Physics and Astronomy Research Council (PPARC).
CPH gratefully acknowledges support from a PPARC studentship.  The Konus-Wind experiment is
supported by Russian Space Agency contract and RFBR grant 06-02-16070.  HAK is also grateful for useful discussions with T. Sakamoto, P. O'Brien and Y. Kaneko.

\clearpage

\begin{deluxetable}{cccc}
\tabletypesize{\scriptsize}
\tablecaption{UVOT Limiting magnitudes\label{tbl-1}}
\tablewidth{0pt}
\tablehead{
\colhead{Filter} & \colhead{Exposure (s)} & \colhead{T$_{mid}$ (s)} & \colhead{3-sigma limit}}
\startdata
V	& 168	& 424	& 19.00\\
B	& 75	 & 524	& 19.59\\
U	& 78	 & 511	& 19.34\\
UVW1	& 78	&  498	& 18.62\\
UVM2	& 78	& 483	& 18.79\\
UVW2	& 68	& 498	& 18.73\\
\enddata
\tablecomments{Data taken from GCN 3638 (Blustin et al).  T$_{mid}$ is the mid-point of the summed observation measured with respect to the BAT trigger time T$_0$.}
\end{deluxetable}

\clearpage



\begin{figure}\caption{Background subtracted BAT light curves, power-law fit indices and hardness ratios for GRB~050717.  The panels on the right show the full duration of the prompt emission; those on the left zoom in to show the precursor peaks in the light curves more clearly.  {\bf Light curves (top four sets of plots): } The rate is corrected for the effective area as a function of source location in the field of view before and during the slew.  After the slew the source is on-axis. The start and end of the slew to the target are shown by vertical lines. The burst duration measures $T_{90}$\ and $T_{50}$\ are shown by horizontal lines in the right hand plots, with $T_{90}$\ shown above $T_{50}$. The time binning is 1~s for the right-hand plots and 64~ms for those on the left. {\bf Power Law fit photon index (fifth set of plots):} Separate fits were made to each time interval indicated.  The BAT data (plain symbols) are best fit by a simple power law.  The plot also shows joint fits to the BAT and Wind data (open diamonds) and to the BAT and XRT data (open square).  For the leftmost  BAT/Wind point, the index $\alpha$\ of the cut-off power law fit (see text) is shown.  For the other joint fit points, the photon index from a power-law fit is shown.  {\bf BAT hardness ratios (lower two sets of plots):} Two sets of ratios (defined on the plot) are shown to illustrate the spectral hardening during the rise to the main peak, followed by a softening as the prompt emission evolves.  The final data points show a second hardening of the spectrum.  The time scale is the same for all plots in a vertical column.\label{fig1}}
\end{figure}
\clearpage
\epsscale{.8}
\plotone{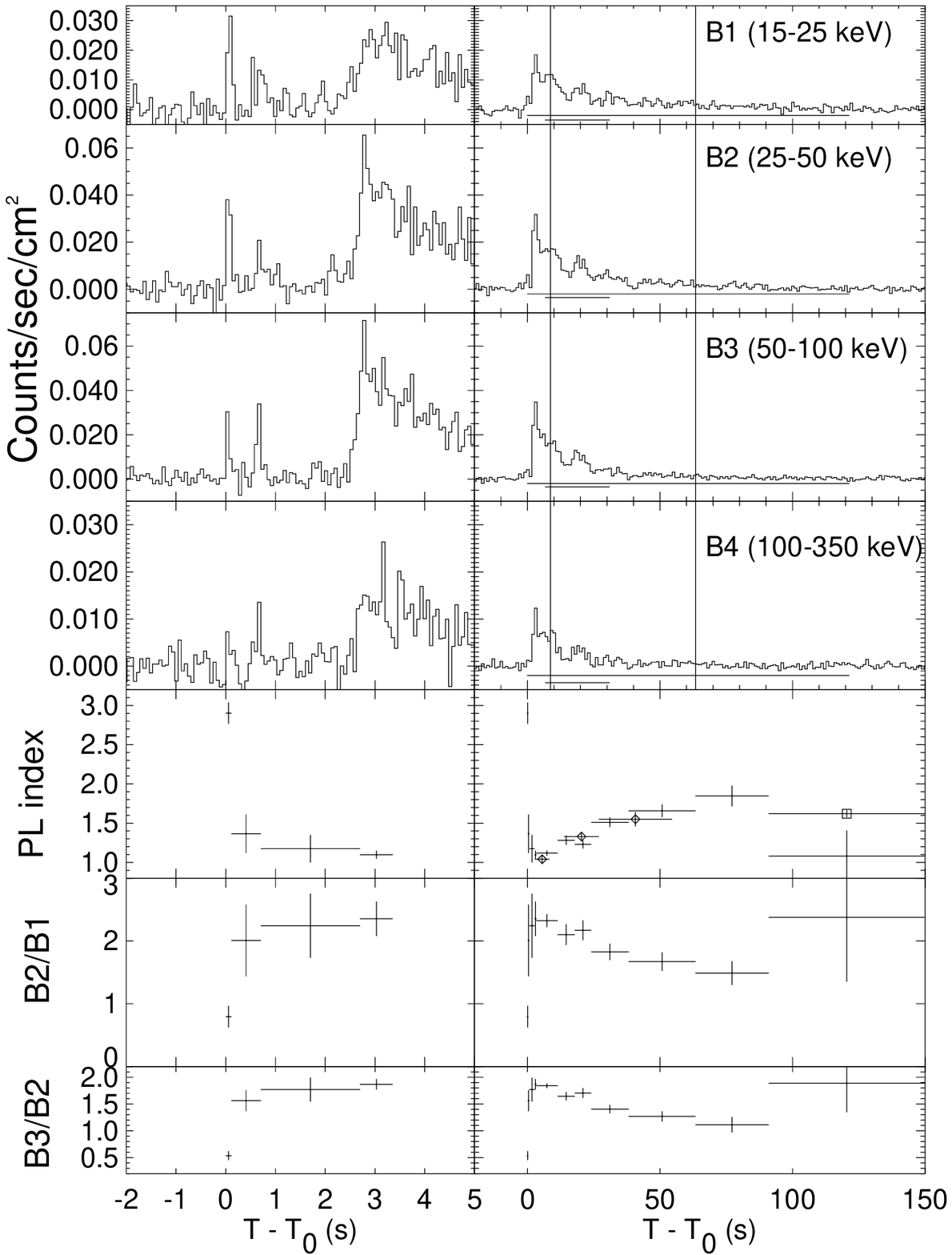}
\epsscale{1}
\centerline{Fig. 1. ---}
\clearpage


\begin{figure}
\plotone{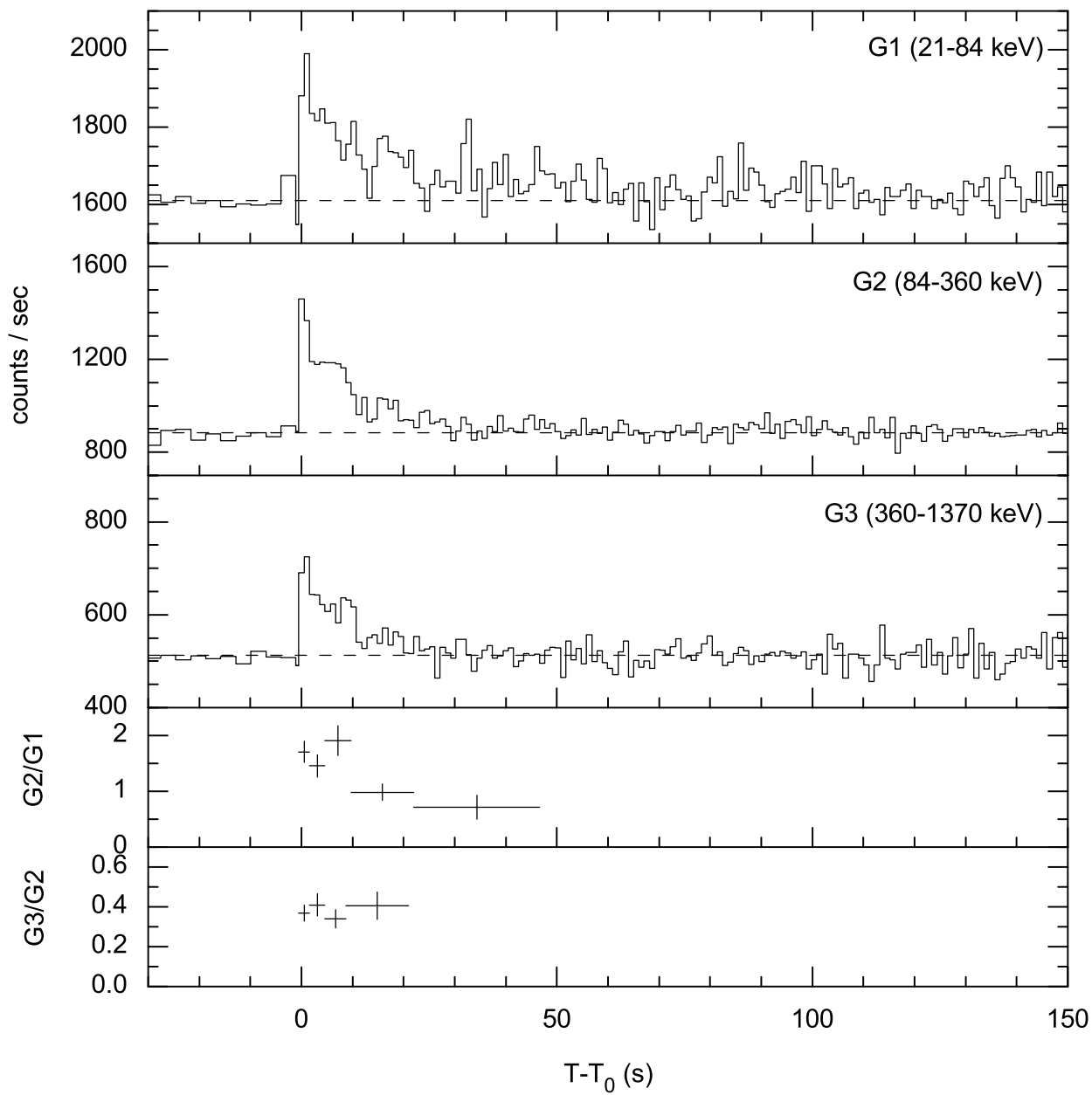}
\caption{The Konus-WIND light curve for GRB 050717 in three energy bands.  
The data before T-T$_0$(K-W) =-0.512~s were recorded in the waiting
mode with 2.944-s time resolution, after that data were recorded at finer time 
resolution and binned at 1.024 sec.  The energy bands used in the hardness ratios at the bottom of the plot are defined in the top panels of the plot.\label{fig2}}
\end{figure}

\begin{figure}
\plotone{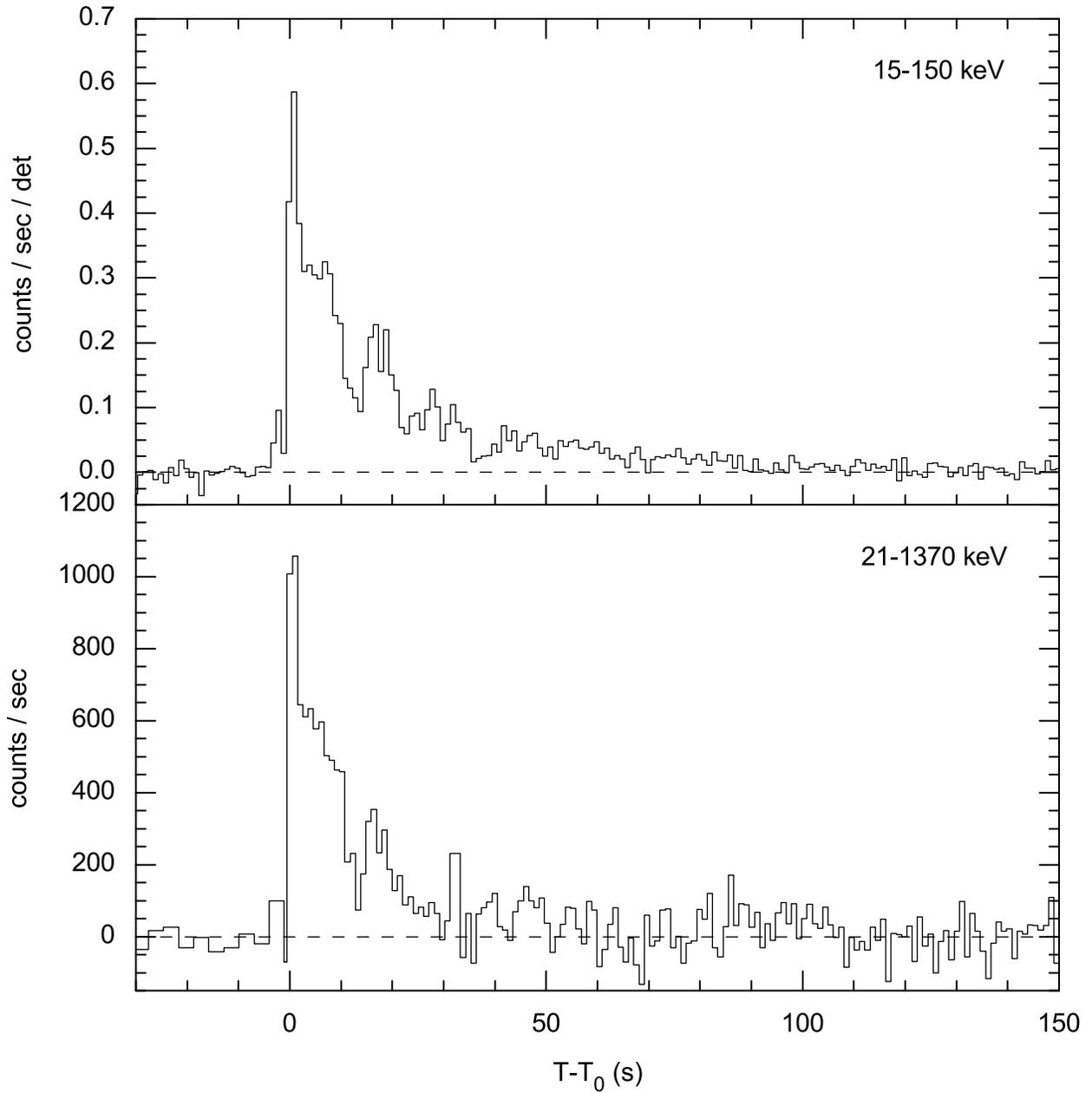}
\caption{The background subtracted BAT (top panel) and Konus-WIND (bottom) light curves on the same time scale.  The plots have been adjusted so that the trigger time for both plots are the same relative to the burst.  This means that T$_0$\ in the lower plot is actually T$_0$(BAT) plus the propagation time between the spacecrafts (2.369 s).\label{fig2a}}
\end{figure}

\begin{figure}
\includegraphics*{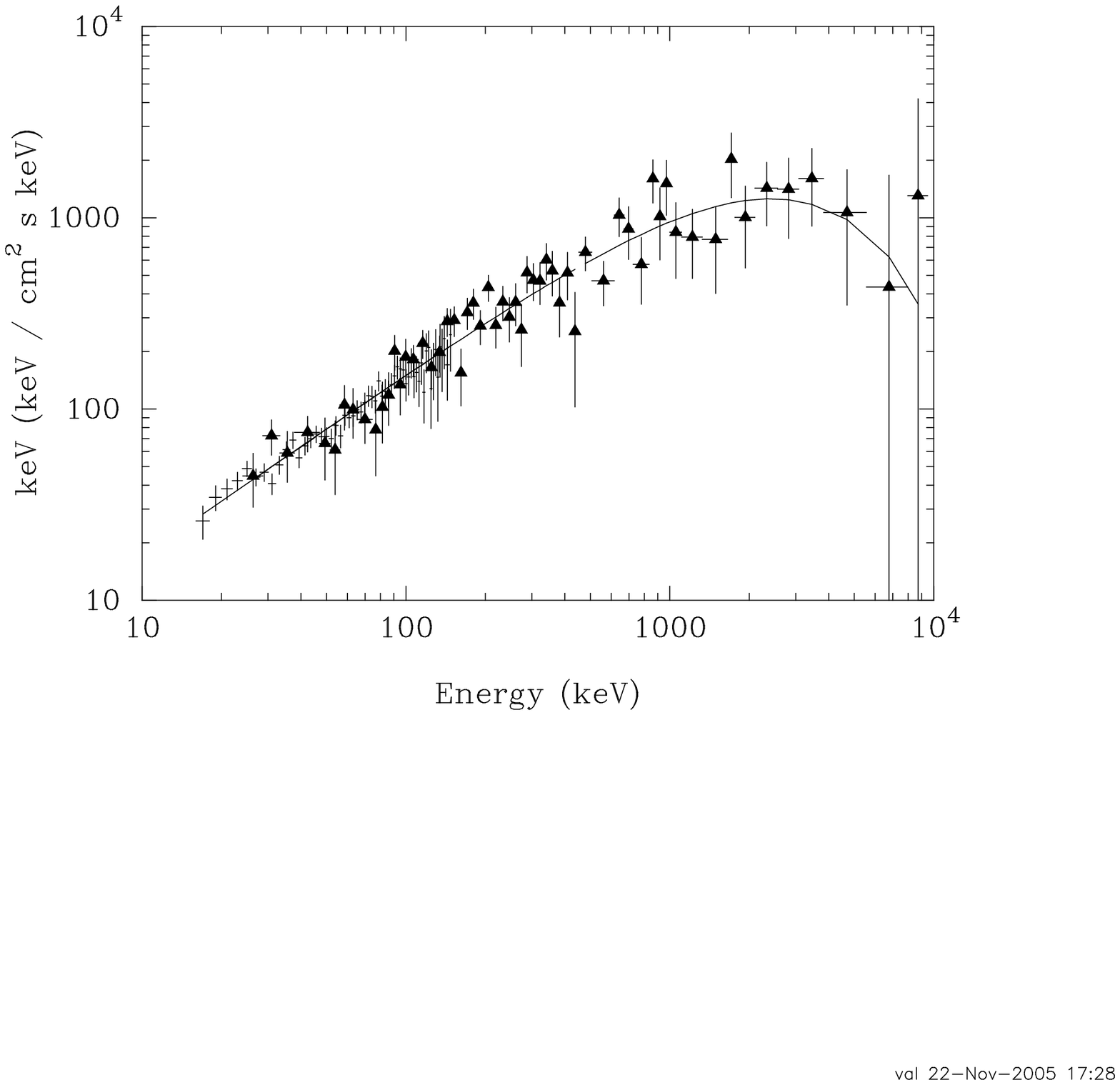}
\caption{Joint fit to a cut-off power law model (defined in the text) for the BAT and Konus-Wind data during the main peak of emission T$_0$+2.843 s to T$_0$+8.219 s.  The  value of $E_{peak}$ for this fit is $2401^{+781}_{-568}$ keV. Points from the BAT spectrum are shown as crosses, those from the Konus spectrum are shown as filled triangles.\label{fig1a}}
\end{figure}

\begin{figure}
\plotone{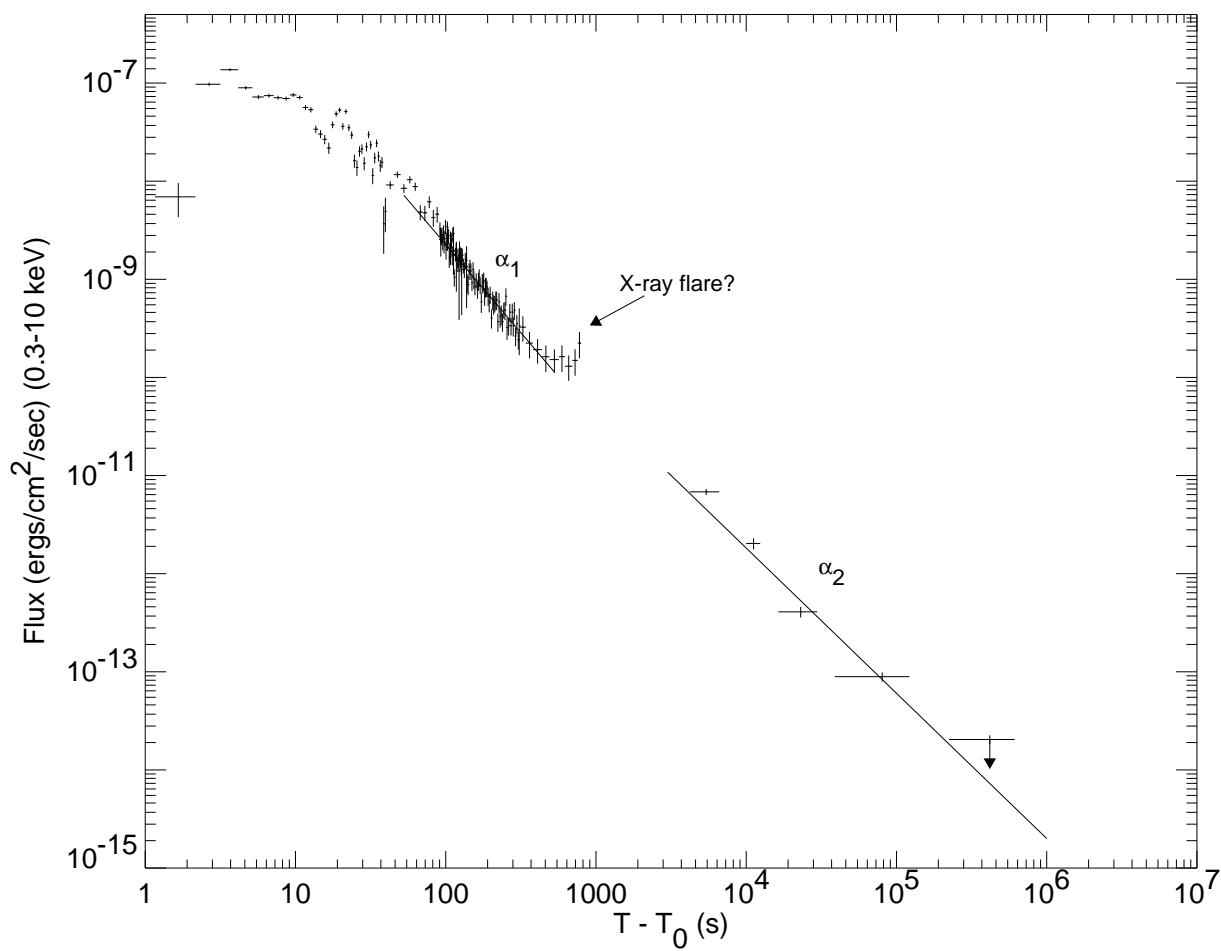}
\caption{The combined BAT prompt emission and XRT afterglow light curve.  Points in the BAT light curve have been extrapolated from the BAT 15--150 keV energy band to the XRT 0.3--10 keV band and corrected for differences in the effective area (see discussion in the text). This shows how the prompt emission makes a smooth transition into the afterglow.  The broken power law fit to the X-ray light curve decay is also shown ($\alpha_1 = 2.10; \alpha_2 = 1.48$). The last data point (upper limit) was combined from five orbits in PC mode.\label{fig3}}
\end{figure}

\begin{figure}
\plotone{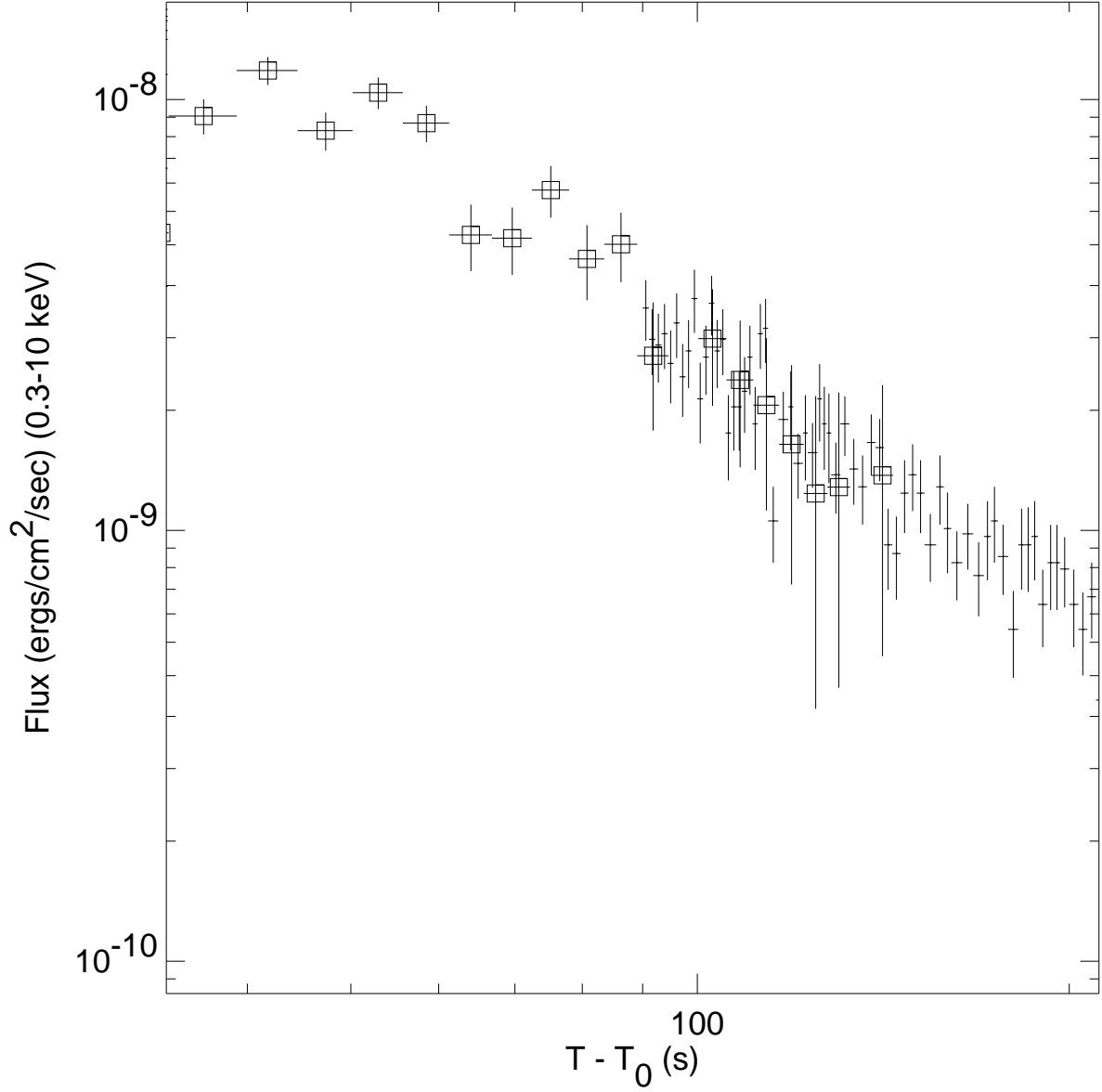}
\caption{The section of Figure~\ref{fig3} showing the overlap between BAT (open squares) and XRT (crosses) emission. This figure clearly shows the smooth transition from prompt gamma-ray to early X-ray emission.  See caption to Figure~\ref{fig3} and the text for a discussion of the extrapolation of the BAT data points.\label{fig4}}
\end{figure}




\end{document}